\documentclass[aps,shownopacs,superscriptaddress,nofootinbib,preprint,11pt]{revtex4}
\usepackage{hyperref}
\usepackage{amsmath}
\usepackage{array}
\newcolumntype{L}[1]{>{\raggedright\let\newline\\\arraybackslash\hspace{0pt}}m{#1}}
\newcolumntype{C}[1]{>{\centering\let\newline\\\arraybackslash\hspace{0pt}}m{#1}}
\newcolumntype{R}[1]{>{\raggedleft\let\newline\\\arraybackslash\hspace{0pt}}m{#1}}
\usepackage{float}
\usepackage{graphics}
\usepackage{graphicx}
\usepackage{epsfig}
\usepackage{psfrag}
\usepackage{color}
\usepackage{slashed}

\usepackage{amsfonts}
\usepackage{amssymb}

\newcommand*{\be}{\begin{equation}}
\newcommand*{\ee}{\end{equation}}
\newcommand*{\bea}{\begin{eqnarray}}
\newcommand*{\eea}{\end{eqnarray}}

\newcommand{\comment}[1]{}

\newcommand{\cref}[1]{Chapter~\ref{c.#1}}

\def\nn{\nonumber \\}

\def\beq{\begin{equation}}
\def\eeq{\end{equation}}
\def\bea{\begin{eqnarray}}
\def\eea{\end{eqnarray}}
\def\ba{\begin{array}}
\def\ea{\end{array}}
\def\bi{\begin{itemize}}
\def\ei{\end{itemize}}
\def\be{\begin{enumerate}}
\def\ee{\end{enumerate}}
\def\bc{\begin{center}}
\def\ec{\end{center}}
\def\bt{\begin{table}}
\def\et{\end{table}}
\def\btb{\begin{tabular}}
\def\etb{\end{tabular}}

\def\lsim{\raise0.3ex\hbox{$\;<$\kern-0.75em\raise-1.1ex\hbox{$\sim\;$}}}
\def\gsim{\raise0.3ex\hbox{$\;>$\kern-0.75em\raise-1.1ex\hbox{$\sim\;$}}}

\begin{document}

\title{Gravitational rescue of minimal gauge mediation}
\author{Abhishek M. Iyer}
\email{abhishek@theory.tifr.res.in}
\address{Department of Theoretical Physics, Tata Institute of Fundamental Research, Homi Bhabha Road, Colaba, Mumbai 400 005, India}
\address{Centre for High Energy Physics, Indian Institute of Science,
Bangalore 560012, India},

\author{ V Suryanarayana Mummidi}
\email{soori@cts.iisc.ernet.in}
\author{Sudhir K. Vempati}
\email{vempati@cts.iisc.ernet.in}
\address{Centre for High Energy Physics, Indian Institute of Science,
Bangalore 560012, India}

\begin{abstract}

Gravity mediated supersymmetry breaking becomes comparable to gauge mediated supersymmetry breaking contributions when messenger masses are close to the GUT 
scale. By suitably arranging the gravity contributions one can modify the soft supersymmetry breaking sector to generate a large 
stop mixing parameter and a light Higgs mass of 125 GeV. In this kind of 
hybrid models, however the nice features of gauge mediation like 
flavour conservation etc., are lost.  To preserve the nice features, 
gravitational contributions should become important for lighter 
messenger masses and should be important only for certain fields. 
This is possible when the hidden sector contains multiple (at least two)
spurions with hierarchical vacuum expectation values. In this case, the
gravitational contributions can be organised to be `just right'. We present
a complete model with two spurion hidden sector where the gravitational
contribution is from a warped flavour model in a Randall-Sundrum setting. 
Along the way, we present simple expressions to handle renormalisation
group equations when supersymmetry is broken by two different sectors
at two different scales.

\end{abstract}
\vskip .5 true cm

\maketitle
\section{Introduction and Motivation}
The discovery of the Higgs boson with a mass $\sim 125~ \text{GeV}$ a couple of years ago \cite{cms1,atlas1,cms2,atlas2} has led to tremendous
excitement in the field. While the discovery has validated the Higgs mechanism of the Standard Model and is broadly consistent with expectations based
on Minimal Supersymmetric Standard Model (MSSM) \footnote{We mean that the observed Higgs is within theoretical upper bound of 135 GeV \cite{Ellis:1990nz,Haber:1990aw,Okada:1990vk}.}, minimal models of supersymmetry breaking like gauge mediation have been strongly constrained by this discovery. 

Gauge Mediation has several nice features that make it an attractive mechanism for supersymmetry breaking : (i) no additional flavour violation in the
soft sector (ii) minimal models have very few parameters, some times as low as one, making them very predictive (iii) different phenomenology compared
to the traditional gravity mediated models. However,  the discovery of the Higgs  boson with  mass $\sim $ 125 GeV,  has strongly constrained these models. 
The one loop corrected  mass for the lightest CP even Higgs boson has the form \cite{Ellis:1990nz,Haber:1990aw,Okada:1990vk,Haber:1996fp,Hahn:2013ria,Buchmueller:2013psa} 
\begin{eqnarray}
m_h^2&=& M_Z^2\cos{2\beta}^2+ \frac{3m_t^4}{4\pi^2 v^2} \left[\log\left(\frac{M_S^2}{m_t^2}\right)+\frac{X_t^2}{M_S^2}\left(1-\frac{X_t^2}{12 M_S^2}\right)\right],
\label{hformula}
\end{eqnarray}
Where $M_S=\sqrt {m_{\tilde t_1}m_{\tilde t_2}}$ and $X_t=A_t-\mu \cot \beta$. For $M_S$ around 1 TeV,  it can be seen that $X_t$ should be close to its
maximal value 
($\approx \sqrt{6} M_S$) to obtain a Higgs mass around 125 GeV \cite{Casas:1994us,Carena:1995bx,Arbey:2011ab,Arbey:2012dq}. It is well known,  in minimal gauge 
mediation, that tri-linear A terms are highly suppressed  at the mediation scale leading to much smaller values of $X_t$ than required at the weak scale. One can of course try to increase the mediation scale to increase the renormalisation group (RG) running from traditional values of $\sim 100$ TeV all the way up to the GUT scale. However it has been shown that a sufficient appreciation in the $X_t$ value is only possible, for $\sim 1$ TeV stops, with heavy gluinos and/or tachyonic stops at high scale \cite{Draper:2011aa}. 
Since $X_t$ is very small  in GMSB, $M_S\sim 4$ TeV is required to achieve the desired Higgs mass. This scenario is not very attractive as it comes at the expense of making the
stop mass eigenvalues out of reach of the LHC. Thus minimal versions of GMSB models would fail to accommodate 125 GeV Higgs, if the SUSY particle masses lie below 2 TeV or so.

Various extensions have been proposed in the literature to make models of gauge mediation compatible with the 125 GeV Higgs mass \cite{Perez:2012mj,Endo:2012rd,Martin:2012dg,Jelinski:2011xe,Yanagida:2012ef,Abdullah:2012tq,Evans:2012hg,Albaid:2012qk,Frank:2013yta,Evans:2013kxa,Craig:2013wga,Calibbi:2013mka, Byakti:2013ti,Evans:2011bea,Craig:2012xp,  Fischler:2013tva, Krauss:2013jva}. 
For instance, one could consider explicit Yukawa like messenger matter mixing in addition to gauge 
interactions\cite{Chacko:2001km,Chacko:2002et,Shadmi:2011hs}. In some cases of this type, it is possible to get solutions with minimal fine tuning 
leading to a naturally light Higgs boson with mass 125 GeV\cite{Evans:2013kxa,davidtalk,Ding:2013pya}. 
Another possibility is to consider, anomaly free, U(1) gauge extension of MSSM gauge 
group\cite{Mummidi:2013hba,Lee:2007fw,Barger:2006rd,Gunion:2009zz,Demir:2005ti,Ma:2002tc,Han:2004yd,Cohen:2008ni,Anastasopoulos:2008jt}.

Alternatively, one can ask the question whether gravity mediated supersymmetry  breaking contribution can help in generating large $A_t$ in GMSB models, while keeping the spectrum within the reach of LHC ($\lesssim$ 2-3 TeV). Note that these gravity contributions are always present in any gauge mediation model as Planck scale
suppressed operators. Typically, this contribution is suppressed due to the relatively `low' values of $F$ -terms required by gauge mediation. For example, the stop tri-linear coupling at the reduced Planck would
would be generated as 
\begin{eqnarray}
A_t &\simeq& \langle F_X \rangle \over M_{Pl}
\nonumber 
\end{eqnarray}
where $ \langle F_X \rangle$, is the vacuum expectation value (vev) of the F-term of the spurion multiplet and $M_{Pl}\sim 10^{18}$ GeV, is the reduced Planck scale.  In gauge mediation models, typically $ \langle F_X \rangle$ lies in the range $\sim 10^{10}  - 10^{12} ~ \text{GeV}^2$. 
This $F_X$ value is too small to have any meaningful impact on the soft sector when contributions from gravity mediation also are included along with GMSB contributions; they  are of the order $\sim (10^{-8}-10^{-6} )$ GeV.  As will be discussed in the next section, if one increases the mass of the messengers the gravitational contribution become increasingly important. This result holds true as long as supersymmetry breaking by as single spurion parametrises the hidden sector.

In this paper, we consider the  case where the  hidden sector is parametrised by more than one spurion fields and show that the gravity contribution to the A-terms could be much larger as one can effectively `decouple' both the gravity and gauge contributions.  In such case, even if the messenger masses are smaller, gravity mediation can play an important role.  Models with multiple spurion fields are quite common,  for example in string based models \cite{Brignole:1997dp,Intriligator:2007py,Intriligator:2007cp,Abel:2006qt,Gaillard:2009rma,Buican:2009zta,Braun:2013wr,Cvetic:2012kj,Goodsell:2010ie,deAlwis:2008kt}, mirage mediation \cite{Everett:2008qy,Choi:2004sx,Choi:2005ge,LoaizaBrito:2005fa,Lebedev:2005ge,Pierce:2006cf,Abe:2006xp,Choi:2006xb} 
and models with multiple hidden sectors \cite{Intriligator:2007py,Intriligator:2007cp,McCullough:2010wf,Craig:2010yf,Choi:2006xb}.  
In the recent times, supersymmetry breaking in multiple sectors has been receiving attention in literature. Most of the works are concentrated on Goldstini sector and their corresponding collider phenomenology \cite{Ferretti:2013wya,Argurio:2011gu,Cheung:2011jq,Cheung:2010mc}.

The combination of  gauge and gravity mediations,  some times called  ``hybrid models'' of supersymmetry breaking
\cite{Dudas:2008qf,Hiller:2008sv} has been studied in the literature with focus on specific issues in phenomenology like flavour or dark matter.  In the present work, we show that this idea can be used  to address the light CP even Higgs mass problem  in minimal gauge mediation.  However, it should be noted that the framework presented here is 
quite different compared to the ones in the literature. In the present work, we focus on the Higgs sector, especially concentrate on how to increase its mass through additional gravity mediation contributions to the minimal messenger model. It turns out that the spurions should have hierarchical F -terms, and furthermore the gravity contributions should be `flavoured' in the sense that the $A$ terms of the third generation should be larger compared to the first two generations. Similarly the soft terms should be such that the total spectrum should still be dominated  minimal gauge mediation. 
 
The paper is organized as follows. In the next section, we discuss the character  of the gravitational contributions required to generate large A-terms in minimal messenger model. We analyze the gravity contributions in GMSB models in both the cases with one and two spurions in the hidden sector. We also study this scenario using semi-analytical expressions for RGs. In section 3, we present an explicit model for gravity sector which has the essential features
and also present explicit numerical examples. We close with a summary. 

\section{Character of Gravitational contributions}
In minimal gauge mediation, soft terms are generated through 1-loop and 2-loop diagrams
involving messenger fields coupled to the MSSM fields through gauge interactions. The messengers
are connected through the hidden sector which is parametrised by a spurion field $X$ as
\begin{equation}
W_{\text{mess}} = \lambda X \Phi \bar{\Phi}
\end{equation} 
where $\Phi, \bar{\Phi}$ are messenger fields and $X = M_X + \theta^2 F_X$ parametrises
the hidden sector. $M_X$ is the mass of the messenger fields and $F_X$ is the supersymmetry
breaking F-term.
The soft terms due to gauge mediation at the messenger scale $M_X$ are given by \cite{Dine:1981gu,Nappi:1982hm,AlvarezGaume:1981wy,Giudice:1998bp}: 
\begin{eqnarray}
M'_a &\approx& \frac{\alpha_a}{4 \pi} \left({F_X\over M_X}\right)  \nonumber \\
m^{\prime2} &\approx& 2  \sum_{a=1}^3 \left(\frac{\alpha_a}{4\pi}\right)^2 C_a \left({F_X^2\over M_X^2}\right)\nn
A'&\approx& 0
\label{gaugesoftterms}
\end{eqnarray}

Typically $\Lambda \equiv F_X/M_X$ is chosen to be around 100 TeV to get low energy soft terms of $\mathcal{O}$(1) TeV. Generally $M_X$ is chosen to be close to $\Lambda$ (twice of $\Lambda$ is the usual choice) 
and it can be as large as the GUT scale. $F_X$ is scaled accordingly while keeping $\Lambda$ fixed. 
The low energy spectrum  is computed by running of the renormalisation group equations (RGE) 
from $M_X$  to the weak scale.  Since we are concerned mostly with the light CP even Higgs mass,
we will concentrate on the parameters relevant for it.  At the weak scale,  the relevant parameters
appearing in eq.(\ref{gaugesoftterms}) can be written as 
 \begin{eqnarray}
m_{Q_3}^2 (M_{weak}) &\approx& 1\times10^{-4}\Lambda^2~\text{GeV$^2$}\nonumber \\
 m_{U_3}^2 (M_{weak}) &\approx& 7.62\times10^{-5}\Lambda^2~\text{GeV$^2$}\nonumber\\
A_t  (M_{weak})  &\approx & -0.002~\Lambda ~\text{GeV}
\label{gaugeanalytic}
\end{eqnarray}
where $\Lambda=\frac{\langle F_X \rangle}{M_X}$. The expressions are evaluated for $M_X=10^{6}$. Keeping $\Lambda$ fixed,
if $M_X$ is increased from $10^6$ to $10^{14}$, then the magnitude of $A_t$ only increases from $0.002\Lambda$ to $0.005\Lambda$. Increasing $\Lambda$
 will also increase $M_{S}$, the SUSY scale, thus spoiling naturalness. It can be shown that even by choosing a large $M_X$ close to the GUT scale, so 
as to increase the logarithmic running, the Higgs mass barely manages to go beyond the LEP
limit of 114.2 GeV from eq.(\ref{hformula}). This is because $X_t$ remains small, as long as $M_S$ is fixed to be 1 TeV. This result is now well known \cite{Draper:2011aa}. 

Gravity mediation will always contribute to the soft terms at the high scale. However, for the full
range of $F_X ~\sim 10^{12}- 10^{21} ~\text{GeV}^2$ (fixing $\Lambda$ = 100 TeV). Gravitational contributions are only significant when the messenger scale reaches the Planck scale. The gravity contributions at the Planck scale are given in terms of effective
operators as 
\begin{eqnarray}
 m_{ij}^2&=&\kappa_{ij}\int d^2\theta d^2\bar{\theta}\left({X^\dagger X \over M_{Pl}^2}\right)\Phi^\dagger \Phi\nn
 M_{ab}&=& f_{ab}\int d^2\theta\left({X \over M_{Pl}}\right)\mathcal {W}^a\mathcal{W}^b\nn
 A_{ijk}&=&\eta_{ijk}\int d^2\theta\left({X \over M_{Pl}}\right)\Phi_i \Phi_j \Phi_k\nn
\label{singlescalegravity}
\end{eqnarray}
where $k_{ij}$ and $\eta_{ijk}$ are arbitrary $\mathcal{O}(1)$ parameters and $f_{ab}$ is the gauge
kinetic function.    
At the messenger scale $M_{X}$, if the running is not significantly large from $M_{Pl}$ to $M_{X}$, 
the soft terms including both gauge and gravity contributions can be parametrised as
\begin{eqnarray}
 \tilde m_{ij}^2 (M_{X}) &\approx & m_{3/2}^2 \left[ {\alpha_a^2 \over (4 \pi)^2} C_ a { M_{Pl}^2 \over M_X^2}~ \delta_{ij} ~+ 
\tilde{\kappa}_{ij}  ~+ ~\beta_{ij} \right] \nn 
\tilde M_{ab} (M_X) &\approx & m_{3/2} \left[ {\alpha_a \over (4 \pi) } {M_{Pl} \over M_{X}} + \gamma_a \right] \delta_{ab} \nn 
\tilde A_{ijk} (M_X) &\approx & m_{3/2} \left[  \tilde{\eta}_{ijk} + \xi_{ijk} \right]  
\label{effectivehybrid} 
\end{eqnarray}
where we have used $\langle F_X \rangle = m_{3/2} M_{Pl}$ and $\tilde{\kappa}$ and $\tilde{\eta}$ are the renormalisation group
corrected $\kappa$ and $\eta$ couplings respectively and are also of $\mathcal{O}$(1)\footnote{Where $\tilde \kappa\approx \kappa$ and $\tilde\eta\approx\eta$.}. $\beta$, $\xi$ and  $\gamma$ are corrections to the original
couplings due to RG running. These parameters contain the Clebsch factors (which could be
large in the presence of GUT group between $M_{Pl}$ and $M_X$) and the logarithmic factors
due to running. Numerically $\beta=\xi\approx \frac{1}{16\pi^2}\text{Log}\left(\frac{M_{Pl}}{10^{16}}\right)\sim0.02$.
At one loop, the coefficients $\gamma$ can be determined exactly and are given as $\gamma_1=0.96,\;\;\gamma_2=0.99,\;\;\gamma_3=1.01$.
We note here that the gravitino mass in this case ($m_{3/2}\sim 100$ GeV) is large as compared to low scale GMSB, where typically $\langle F_X\rangle << M_{Pl}^2$.
This is because, keeping $\Lambda=10^5$ GeV fixed, a choice of a heavy messenger scale $M_X\sim 10^{16}$ GeV also pushes the vacuum expectation value of the spurion to $\langle F_X\rangle \sim 10^{21}$ GeV$^2$.
This framework is similar to the one presented in Ref. \cite{Dudas:2008qf,Dudas:2008eq}. 
There it has been shown that an explicit realization requires certain conditions on the messenger sector.

From the above effective parametrisation, it is clear that for heavy messengers with mass $\sim M_{Pl}$ scale, the gravitational mediation contributions can become comparable and can significantly alter the spectrum. In the effective picture represented by Eq.(\ref{effectivehybrid}), we can choose the gravity coefficients such that soft terms to have the required form to give the correct Higgs mass at the weak scale. For example, using the analysis of \cite{Draper:2011aa}, we can choose the gravity contributions to the stop sector and gluino sector such that they generate a large enough $A_t$ at the weak scale after RG running.   

From the above discussion, one can conclude that gravitational contributions can indeed rescue minimal gauge mediation if the messenger masses are sufficiently high and they are of the right type. The question then arises what happens if the messenger masses are not so heavy and are only of the order of 100 TeV or so; Can gravitational contributions still play a role ? This is the question we try to address now.  

The advantage of such a `low' scale for $M_X$ is that the nice  features of gauge mediation are left intact.  
However,  gravitational mediation contributions to the soft sector are small as discussed above. One simple
way to increase the gravitational mediation would be to consider two different sectors of supersymmetry
breaking.  While only one of them couples to the messengers, both couple to gravity.  In this set up, 
let us try to understand the character of the gravitational contributions.  

We will consider two spurions $X_1$ and $X_2$ parametrising the two sectors where supersymmetry is
broken. As before we denote $X_1 \equiv M_{X_1} + \theta^2 F_1$, which couples to the messengers at the 
scale $M_{X_1} ~\sim~100$~ TeV.  The other  spurion $X_2$ which parametrises the sector which does not
couple to messengers \footnote{In the limit both these sectors are sequestered, we will have a new particle in the spectrum, the Goldstino, which we will not consider here \cite{Ferretti:2013wya,Argurio:2011gu,Cheung:2011jq,Cheung:2010mc}.}. In terms of the
effective operators, just below $M_{Pl}$ scale, the following  are the contributions from the gravitational mediation:

\begin{eqnarray}
 m_{ij}^2&=&\int d^2\theta d^2\bar{\theta}    \left(   k_{ij}  {X_1^\dagger X_1 \over M_{Pl}^2}  
+  k'_{ij}  {X_2^\dagger X_2 \over M_{Pl}^2}   \right) \Phi^\dagger \Phi\nn
M_{ab}&=& f_{ab}\int d^2\theta\left({X_1 \over M_{Pl}}  + {X_2 \over M_{Pl}} \right)\mathcal {W}^a\mathcal{W}^b\nn
A_{ijk}&=&\int d^2\theta\left(\eta_{ijk}   {X_1 \over M_{Pl}}  +  \eta'_{ijk}   {X_2 \over M_{Pl}} \right)\Phi_i \Phi_j \Phi_k\nn
\label{doublescalegravity}
\end{eqnarray}
where $F_1 \ll F_2 $.  At the scale $M_{X_1}$ additional contributions from gauge mediation have the same form 
as in Eq.(\ref{gaugesoftterms}) with $F$ replaced by $F_1$. To obtain the weak scale spectrum, we use semi-analytical expressions given in Appendix~\ref{appendix1}. We observed that in this case, there is a nice way of organising the  weak scale soft terms which are a result of the RGE evolution from $M_{Pl}$ to $M_{X_1}$ and then to $M_{weak}$. 
Through out this paper, we will use primed objects to denote the contributions from gauge mediation sector. 
Unprimed objects will denote those from gravity mediation. The total soft terms have a $\tilde{}$~ on them. 
At the weak scale, we have the following relations: 
\begin{eqnarray}
\tilde{M}_a (weak)&=&M_a(weak)+M'_a(weak) \nn
\tilde{A}(weak)&=&A(weak)+ A'(weak) \nonumber 
\label{hybrid1}
\end{eqnarray}
Thus both the gaugino masses and the tri-linear terms just add up at the weak scale as though they 
have no knowledge about the presence of another supersymmetry breaking sector\footnote{These results are true only up to one loop order. Issues about higher loops and 
multiple (greater than 2) SUSY breaking sectors will be considered elsewhere. }. 

The scalar masses however, as they depend quadratically on the gaugino masses do not add
linearly. They have mixed contributions from both the supersymmetry breaking sectors. Schematically
they can be written as (neglecting the Yukawa contributions) 
\begin{equation}
\tilde{m}_{\tilde{f}}^2  (weak) = m_{\tilde f }^{\prime 2} (weak) +  m_{\tilde f}^ 2 (weak) + \sum_a M_a M'_a \zeta_a \zeta'_a 
\label{softhybrid}
\end{equation}
$\zeta(\zeta')$ are the RG factors from $M_{X_2}=M_{Pl}$ to weak scale and $M_{X_1}$ to weak scale
respectively (for more details, see the Appendix~\ref{appendix1} ). 

\begin{figure}[here]
\centering
\includegraphics[width=9cm]{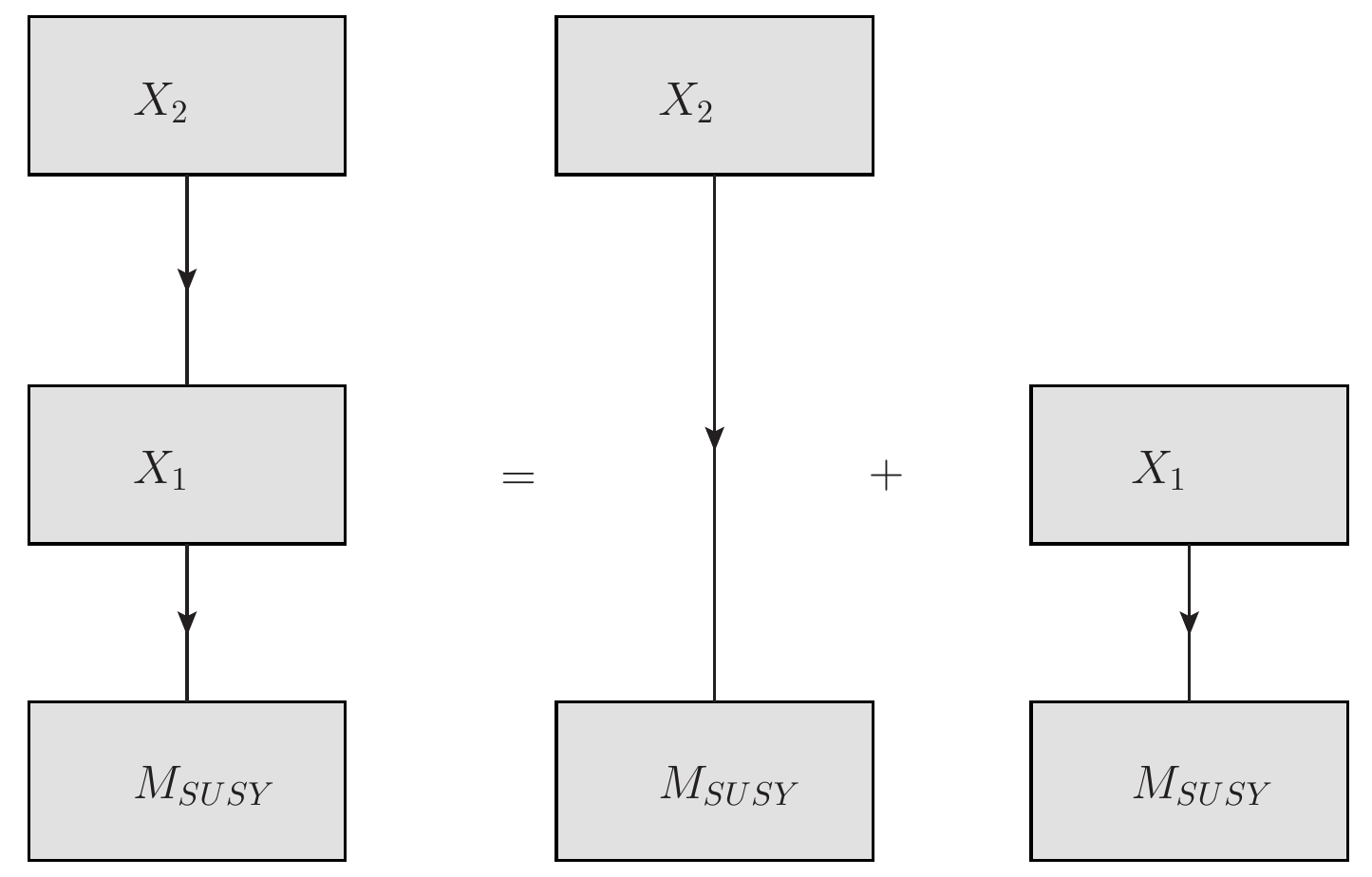}
\caption{RGE running of gauginos and A terms in hybrid SUSY breaking}
\label{rge1}
\end{figure}

\begin{figure}[here]
\centering
\includegraphics[width=12cm]{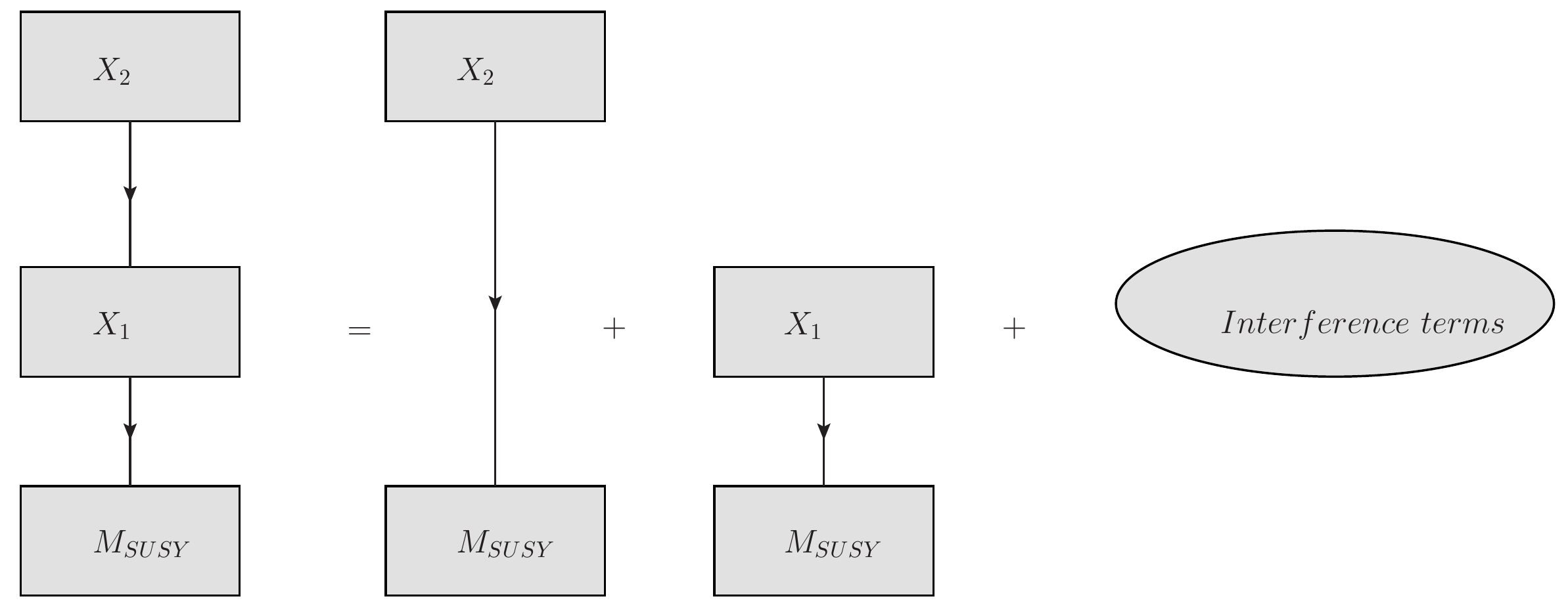}
\caption{RGE running of scalar squared masses in hybrid SUSY breaking}
\label{rge2}
\end{figure}

The total solution can be schematically represented by the following two figures: Fig.~(\ref{rge1}) represents the total solutions for the gaugino masses $M_i$ and the 
tri-linears where as Fig.~({\ref{rge2}}) represents the sfermion mass squared terms. The total gaugino masses and the A terms, at the weak scale, are just the linear sums of the weak scale solutions of the two independent sectors 
which is described in Fig.~(\ref{rge1}). The scalar mass terms at the weak scale, however, contain cross terms mixing contributions from both of the sectors which is described in Fig.~(\ref{rge2}). 
The solution of the RGE presented here can be generalised easily for any scenario where supersymmetry
is broken at two different scales and preserves the MSSM structure. Similarly it can be generalised to 
more than two sectors of supersymmetry breaking at different scales. 

In Table [\ref{example1}], we give a numerical example of low energy spectrum due to supersymmetry breaking derived from two different hidden sectors \textit{i.e.} gravity
and gauge mediation. The first two rows are the low scale values of the soft masses from SUSY breaking due to gravity and gauge mediation respectively.
The third row gives the low scale values when SUSY breaking due to both gravity and gauge mediation are present at two different scales. 
As expected, we find that the $A$ terms and the gaugino masses, at the weak scale, to be the linear sum of the gauge and gravity contributions individually\footnote{The slight
discrepancy in the $A$ terms due to individual sum of the two contributions and that due to hybrid SUSY breaking which can be attributed to the difference between numerical and analytic
computations. It is independent of the boundary parameters.}.
The soft masses on the other hand do not add up linearly. For illustration, we present the results for third generation squarks. 
The pure interference terms are listed in the fourth row. The interference terms could become important when the contributions to the soft terms due one of the sectors of
SUSY breaking is sub-dominant.

\begin{table}[h]
\renewcommand{\arraystretch}{1.5}
\begin{tabular}{|c|c |c |c |c |c |c|c|}
\hline
&$m_{\tilde Q_3}^2$ $[\text {GeV}^2]$&$m_{\tilde U_3}^2$ $[\text{GeV}^2]$&$m_{\tilde D_3}^2$ $[\text{GeV}^2]$&$A_t$[GeV] & $M_1$[GeV]&$M_2$[GeV]&$M_{3}$[GeV]\\ 
\hline
Gravity mediation& 3.2$\times 10^6$& 2.1$\times 10^6$& 3.5$\times 10^6$&-2091 & 456.5 & 846.9 & 2191.8\\
\hline
Gauge mediation& 1.04$\times 10^6$& 0.84$\times 10^6$& 1.01$\times 10^6$&-280 & 139.0 & 263.7 & 722.94\\
\hline
Hybrid susy breaking& 4.67$\times 10^6$& 3.43$\times 10^6$& 5.04$\times 10^6$&-2283.4 & 603.12 & 1110.09 & 2817.3\\
\hline
Interference terms &0.43$\times 10^6$&0.49$\times 10^6$&0.53$\times 10^6$&$\sim 0$&$\sim 0$&$\sim 0$&$\sim 0$\\
\hline
\end{tabular}
\caption{To illustrate the organizing principle discussed in the text,scalar squared masses, tri-linear coupling $A_t$ and gauginos are presented for different mediation schemes. Rest of the parameters are fixed as $\tan\beta$=10 for all the schemes, For gravity mediation: $m_0$=500 GeV, $M_{1\over 2}$=1000 GeV, $A_0$=-1000 GeV, For gauge mediation:$\Lambda=10^5$ GeV and $M_{X_1}$=$10^6$ GeV}
\label{example1}
\end{table}

Returning to the problem at hand,  we demand that the gravitational contributions in this case should
be ``just right"  to generate the right Higgs mass. This would mean that gravity should contribute 
significantly to $A$ (mainly $A_t$)  terms but not to anything else.  Practically such a situation 
is hard to achieve  unless one assumes a flavoured supersymmetry breaking in the gravity sector. 
Such examples are abound in literature either based on U(1) symmetries \cite{Binetruy:1996uv,Dudas:1995eq}
or extra dimensions \cite{Nomura:2007ap}.  In the following, we present an example based on 
Randall-Sundrum Models \cite{RS}.  

Before concluding this section, an important comment is in order. For the gravitational contributions to be of the same order as the gauge mediated contributions, the condition can be summarised as $ F_{X_1} /F_{X_2} \sim M_{X_1}/M_{X_2}$ \footnote{More accurately it is represented as $\tilde \alpha\, F_{X_1} /M_{X_1} \sim y\,F_{X_2}/M_{X_2}$. Where $\tilde \alpha$ is gauge coupling in GMSB where as y is the appropriate Yukawa coupling in gravity mediation mechanism.}.  A hidden sector 
model which allows for such hierarchical F-term vacuum expectation values (vevs) would be needed to provide such a solution. In the present work, we do not construct explicit hidden sector models with such properties, but however, assume existence of such models.

\section{A complete example with Randall Sundrum  at high scale}
The Randall-Sundrum (RS) setup consists of a single extra-dimension of radius $R$ compactified on a $S_1/Z_2$ orbifold. 
Two branes, IR and UV are located at the $y=\pi R$ and $y=0$ orbifold fixed points respectively.
This model was originally proposed as solution to the hierarchy problem
by means of a geometric warp factor defined  as
\begin{equation}
 \epsilon=e^{-kR\pi}\sim 10^{-16}
\end{equation}
where $k$ is the reduced Planck scale and $kR\sim 11$. The extent of warping is set by the radius $R$ of the extra-dimension. UV sensitivity of the Higgs mass is `warped' down to the weak scale by the relation $M_{IR}=M_{ew}\sim e^{-kR\pi}M_{Pl}$.

Here we consider a modification of the RS setup with a similar background geometry but with a radius $R'\sim R/8$.
As a result the IR scale corresponds to the GUT scale, $M_{GUT}$. Such frameworks have been considered \cite{Marti,choi1,choi2,Dudas,Brummer,Iyer:2013axa}. 
The spectrum of the effective 4D theory below the GUT scale is that of MSSM.

We assume the two Higgs doublet $H_u$ and $H_d$ to be localized on the low energy brane (IR brane). Matter and gauge fields are in the bulk.
Supersymmetry breaking spurion $X_2=\theta^2 F_{X_2}$ is  introduced on the IR brane. SUSY breaking terms are generated by the brane local interactions of the spurion with the bulk fields.
The $F$ term ($F_{X_2}$) of the spurion $X_2$ develops a vacuum expectation value and generates the soft masses.
In the canonical basis, the soft breaking terms are of the form \cite{Marti,choi1,choi2,Dudas}: 
\begin{eqnarray}
 m^2_{H_u,H_d} &=&  \hat\beta_{u,d} ~m_{3/2}^2\nonumber\\
 (m_{\tilde{f}}^2)_{ij} &=& m_{3/2}^2~\hat \beta_{ij}~e^{(1-c_i-c_j)k R \pi}\xi(c_i)\xi(c_j)\nonumber\\
  A^{u,d}_{ij}&=&m_{3/2} A'_{ij}e^{(1-c_{i}-c_{j})kR\pi}\xi(c_i)\xi(c_j)\nonumber\\
  m_{1/2}&=&k_h m_{3/2}
\label{soft}
\end{eqnarray} 
where the gravitino
mass is defined as 
\begin{equation}
m_{3/2}^2 = { \langle F_{eff} \rangle^2 \over k^2 } = { \langle F_{eff} \rangle^2 \over M_{Pl}^2 }
\end{equation}
where $F^2_{eff}=F^2_{X_2}+F^2_{X_1}$
$\hat \beta_{ij},A',k_h$ are dimensionless $\mathcal{O}(1)$ parameters.
$\xi(c_i)$ is defined as 
\begin{equation}
   \xi(c_i) =  \sqrt{\frac{0.5-c_{i} }{e^{(1-2 c_i)\pi k R}-1}},
   \label{xi}
 \end{equation}

From eq.(\ref{soft}), it is clear that the soft masses follow the structure of the fermion masses which are determined by 
\begin{equation}
(m_{{F}})_{ij} =v\,\hat y_{ij}\,e^{(1-c_i-c_j)k R \pi}\xi(c_i)\xi(c_j)
\end{equation}
where $v$ is the vacuum expectation value and $\hat y_{ij}$ are $\mathcal{O}(1)$ Yukawa couplings. The resulting spectrum is flavorful in the super CKM basis. In particular, the stop sector along with $A_t$ is enhanced naturally as it follows the top mass. The other sectors($12$ and $23$) however, are highly suppressed as they follow the CKM matrix.

We now provide a numerical example of this case where we use a set of $c$ parameters to determine the boundary conditions for the soft masses at $M_{GUT}$. 
The results of the analysis of \cite{Iyer:2013axa} are used where the technique of $\chi^2$ minimization was used to determine the $c$ parameters.
A particular choice of $c$ parameters which we use for our analysis is given in Table~[\ref{example}].
\begin{table}
\renewcommand{\arraystretch}{1.3}
\begin{tabular}{|cccc||cccc|}
 \hline
  \multicolumn{4}{|c||}{Hadron}&\multicolumn{4}{|c|}{Lepton}\\
 parameter & Value &parameter & Value& parameter & Value& parameter&Value\\
 \hline
$c_{Q_1}$&1.49&$c_{D_1}$&1.52&$c_{L_1}$&1.59&$c_{E_1}$&1.32\\
$c_{Q_2} $&1.09&$c_{D_2}$&1.68&$c_{L_2}$&1.18&$c_{E_2}$&1.15 \\
$c_{Q_3}$&-1.59&$c_{D_3}$&1.28&$c_{L_3}$&1.37&$c_{E_3}$&-.22  \\ 
$c_{U_1}$& 2.0&$c_{U_2}$&1.89&-&-&-&-\\
$c_{U_3}$&0&-&-&-&-&-&-\\
\hline
\end{tabular}
\caption{Example point of the bulk mass parameters which satisfy both fermion mass fits.}
\label{example}
\end{table}

We use a slightly modified version of SUSEFLAV \cite{Chowdhury:2011zr} to determine the low energy spectrum. 
The boundary conditions for the sfermion masses due gravity mediation are obtained by using the points in Table \ref{example} in Eq.(\ref{soft}). Fig \ref{soft} gives
the GUT scale contributions to the soft masses as a function of $c$ parameters. Since the lighter generations are located away from the IR brane where the Higgs fields and SUSY breaking, their corresponding GUT scale contributions would be very tiny. 

\begin{figure}[here]
	\includegraphics[width=0.5\textwidth,angle=0]{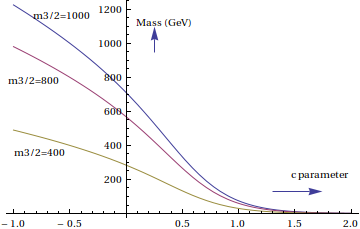}
	\caption{GUT scale contribution to the soft mass parameters as a function of $c$ parameters.}
	\label{soft}
\end{figure}
While there is sufficient choice and freedom among the $\mathcal{O}(1)$ parameters from $1/10$ to 10, below we describe one particular example.
	The gravitino mass is chosen to be $m_{3/2}=\frac{\langle F_2 \rangle}{M_{X_2}}=1000$ GeV. 
	In the slepton mass matrix, the $\mathcal{O}(1)$ parameters in Eq. \ref{soft} is chosen to be  $\hat\beta_{ij}\sim 0.3$.
	For the squark mass matrix we choose $\hat\beta_{ij}= 1$.
	 As the Higgs doublets and the SUSY breaking spurion are localized on the same brane, fitting smaller masses for the fermions would also result in the corresponding slepton masses to the small.
As a result the soft masses for the lighter generations are nearly zero at the Planck scale. As a result of this the right handed sleptons will have a tendency to be the lightest supersymmetric particle (LSP) as can be seen from the following term in its one loop beta-function
\begin{eqnarray}
S=\tilde m^2_{H_u}-\tilde m^2_{H_d}+Tr\left(\tilde m^2_Q-2\tilde m^2_U-m_L^2+\tilde m_D^2+\tilde m_E^2\right)
\end{eqnarray}
As the slepton masses are nearly zero for all three generations the above term is dominated by the contribution due to third generation squarks.
If $S$ is positive, the contributions due to $M_1$ and $S$ go in opposite directions which has a tendency to decrease the slepton mass further. $S$ can be made negative by choosing the Higgsino mass to be non-universal at the high scale. Increasing $\hat M_1$ is not useful as it also increases the mass of the lightest neutralino thereby still retaining the right handed slepton as the LSP. $S$ can be made negative by choosing the Higgsinos to be non-universal, in particular $\tilde m^2_{H_u}<\tilde m^2_{H_d}$. In particular we choose
\begin{eqnarray}
\tilde m^2_{H_u}=500~\text{$GeV^2$}\;\;\;\;\tilde m^2_{H_d}=1000~\text{$GeV^2$}
\end{eqnarray}

 For the electroweakinos, $M_{1,2}=50$ GeV is chosen to lower the contribution to the sleptons. This corresponds to a choice of $k_h=0.07$ in Eq.(\ref{soft}) while we chose  $M_3=500$ GeV for the super-partner mass of the SU(3) gauge field. On account of the small masses of the gauginos the sfermions do not run significantly and the gauge mediation effects will have a tendency to dominate.

For the trilinear couplings we choose $A'_{ij}=Y'_{ij}$ except for $i=j=3$ where $Y'_{ij}$ is the $\mathcal{O}$(1) Yukawa coupling which fits the fermion masses. 
The tri-linear coupling for the top, $A^u_{33}$, is expected to be naturally large compared to those corresponding to the lighter generations.
This can be attributed to the fact that the structure of the tri-linear coupling in this model is exactly similar to the corresponding mass matrices as shown in 
Eq. \ref{soft}. Thus the $c$ parameters which fit a large top mass, will also generate a corresponding larger tri-linear coupling for the top.
The couplings can be further enhanced by choosing the $\mathcal{O}(1)$ parameter
$A'_{33}=-2.5~Y'_{33}$. 
The boundary conditions for gauge mediation are obtained at $M_{X_1}=10^6$ by choosing $\Lambda=2.293\times 10^5$.

Table[\ref{exampleRS}] gives the spectrum for the squarks at $M_S$ in a scenario where two sources of supersymmetry breaking is implemented in a high scale RS model.    
\begin{table}[h]
	\renewcommand{\arraystretch}{1.5}
	\begin{tabular}{|c|c |c |c |c |c |c|c|c|c|c |c |c |c|c|c|}
		\hline
		$\times 10^6$&$m_{\tilde Q_1}^2$ &$m_{\tilde Q_2}^2$ &$m_{\tilde Q_3}^2$ & $m_{\tilde U_1}^2$ &$m_{\tilde U_2}^2$ &$m_{\tilde U_3}^2$& $m_{\tilde D_1}^2$ &$m_{\tilde D_2}^2$ &$m_{\tilde D_3}^2$&$m_{\tilde L_1}^2$ &$m_{\tilde L_2}^2$ &$m_{\tilde L_3}^2$& $m_{\tilde E_1}^2$ &$m_{\tilde E_2}^2$ &$m_{\tilde E_3}^2$ \\ 
		\hline
		Gravity mediation& 1.20&1.20& 1.90&1.27 & 1.27 & -1.33 & 1.20&1.20&1.18&0.03&0.03&0.03&-0.05&0.05&0.005\\
		\hline
		Gauge mediation& 5.45&5.45& 5.04&4.91 & 4.91 & 4.12 & 4.86&4.86&4.84&0.65&0.65&0.65&0.16&0.16&0.16\\
		\hline
		Hybrid susy breaking& 7.1&7.1& 7.23&6.5 & 6.5 & 2.7 & 6.5&6.5&6.94&0.68&0.68&0.68&0.105&0.105&0.16\\
		\hline
		
	\end{tabular}
	\caption{Comparison of contributions to the low energy spectrum for the sfermions due to gravity and gauge mediation incorporated in a high energy RS model for $tan \beta=10$. All numbers are in units $10^6~ \text{GeV}^2$}
	\label{exampleRS}
\end{table}

\begin{table}[h]
	\renewcommand{\arraystretch}{1.5}
	\begin{tabular}{|c|c |c |c |c|}
		\hline
		&$M_1$ &$M_2$ &$M_3$&$A_t$  \\ 
		\hline
		Gravity mediation& 17.74&32.12& 1276.21&-1960.1\\
		\hline
		Gauge mediation& 303.6&564.26& 1459.8&-505.65\\
		\hline
		Hybrid susy breaking& 339.80&624.80& 2568.9&-3075.3\\
		\hline
		
	\end{tabular}
	\caption{Comparison of contributions to the low energy spectrum for the gauginos and $A_t$ due to gravity and gauge mediation incorporated in a high energy RS model for $tan \beta=10$. All numbers are in $\text{GeV}$}
	\label{exampleRS2}
\end{table}

We find that in the absence of contribution to the soft masses at the messenger scale due to gauge mediation, this choice of boundary conditions at the GUT scale leads to a tachyonic spectrum at $M_S$. However once the contributions due to gauge mediation are  on, we get an acceptable low energy spectrum as shown in Table [\ref{spectrum}].

 Table \ref{spectrum} gives the complete low energy spectrum at $M_S$ in the presence of both the sources of SUSY breaking. We see that it is possible to accommodate a heavy Higgs mass of $\sim 125$ GeV in a low scale gauge mediation scenario with fairly light stop masses.
The rest of the spectrum including the gluinos are reasonably heavy owing to contribution to $M_3$ from the gravity sector as well as the gauge sector.
	The dominant contributions are due GMSB as shown in Table \ref{exampleRS2}.
Thus the role played by flavourful gravity mediation is to only generate large trilinear coupling at the gauge messenger scale

\begin{table}[htbp]
\renewcommand{\arraystretch}{1.5}
\begin{center}
\begin{tabular}{|c|c|c|c|c|c|c|c|}
\hline
Parameter&Mass(TeV)&Parameter&Mass(TeV)&Parameter&Mass(TeV)&Parameter&Mass\\  
\hline 

$\tilde t_1$ &   1.79&$\tilde b_1$  &   2.62& $\tilde\tau_1$ & 0.46&$\tilde\nu_\tau$ &  0.81\\
$\tilde t_2$ &   2.74&$\tilde b_2$  &   2.72&$\tilde\tau_2$ &   0.82&$\tilde\nu_\mu$ &  0.81\\
$\tilde c_R$ &   2.61&$\tilde s_R$ &   2.63&$\tilde\mu_R$  &   0.33&$\tilde\nu_e$ &   0.81\\
$\tilde c_L$ &   2.74& $\tilde s_L$ &   2.74&$\tilde\mu_L$  &   0.82&-&-\\
$\tilde u_R$ &   2.61& $\tilde d_R$  &   2.63&$\tilde e_R$   &   0.33&$C_1$&0.61\\
$\tilde u_L$ &   2.74&$\tilde d_L$  &   2.74& $\tilde e_L$   &   0.82&$C_2$&2.01\\
	    $m_{A^0}$&2.69&$m_H^{\pm}$&2.69&$m_h$&0.1223&$m_H$&2.69\\
	    $N_1$&0.33&$N_2$&0.63&$N_3$&2.35&$N_4$&2.35\\
 \hline 
\hline 
\end{tabular}
\end{center}
\caption{Soft spectrum at $M_{S}=2.15$ TeV due to both gravity and gauge mediation contributions. The messenger scale for gauge mediation is chosen to be $M_{X_1}=10^6$.
	 The gluino mass is $m_{\tilde g}=2.6$ TeV and we choose $tan \beta=10$. }
\label{spectrum}
\end{table}

The origin of flavourful soft masses at the GUT scale could potentially lead to large flavour violation at the low scale \footnote{Since the lowest KK masses are $\mathcal{O}(M_{GUT})$
they do not contribute to the flavour processes.}.
The most stringent constraints are due to transitions
between the first two generations in both the leptonic and the hadronic sector. As mentioned earlier, since the lighter generations are localized away from source of SUSY breaking spurion $X$ 
(and the Higgs) the off-diagonal elements as well the diagonal elements are very small at the GUT scale. 
Consider the flavour violating parameter $\delta^f$ defined as
\begin{equation}
 \delta^f_{ij} (i \neq j) = { (U^\dagger \tilde m^2_{f} U )_{ij}  \over  \tilde m^2_{susy}} (i \neq j)
\end{equation}
$U$ is rotation matrix which rotates from flavour basis to the mass basis and $\tilde m^2_{susy}$ is the geometric mean of diagonal elements $\tilde m^2_{ii}$ and $\tilde m^2_{jj}$, $i\neq j$.
 At $M_{GUT}$, the off diagonal elements for the sfermion mass matrices are simply expressed as
\begin{equation}
(\tilde m^2_{f})_{ij}=\tilde m_{ii}\tilde m_{jj}
\end{equation}
where $\tilde m_{jj}=\sqrt{\tilde m^2_{jj}}$.
This follows from the construction in Eq.\ref{soft} and holds when the $\mathcal{O}(1)$ parameter $\hat \beta_{ij}=1$ for all elements of the soft mass matrices. As a result $\delta_{LL,RR} \sim \mathcal{O}(1)$ at $M_{GUT}$. The tri-linear couplings on the other hand are aligned with fermion mass matrices when $A'_{ij}=Y'_{ij}$.
Thus in the SUPER-CKM basis one can expect the tri-linear couplings corresponding to the down sector fermions, $A^{d,E}$, to be nearly proportional to identity.
On the other hand $A^{u}$ will have a hierarchical structure like $V_{CKM}$. As a result the dominant contributions due to FCNC will be due to the sfermion soft mass matrices.

The diagonal elements however, receive significant contributions due to RGE
effects and hence are large at $m_{susy}$. Additionally unlike the off diagonal elements, the diagonal elements also receive contributions due to gauge mediation at the messenger threshold. The off-diagonal elements on the other hand do not evolve much and are small even at $m_{susy}$. Thus, while there could be large flavour
violation at $M_{GUT}$, $\delta^f_{ij}$ reduces from $\mathcal{O}$(1) values at the high scale to values consistent with the experimental bounds at the low scale. Table \ref{deltahigh} gives the high scal $\delta_{ij}$ for both the squark and the slepton sector while Table \ref{delta} gives the corresponding low scale values. We find that while there is potentially large flavour violation at the high scale, the significant running of the diagonal elements push the $\delta$ to be consistent with the contraints given in Table \ref{exptconst}. As a result the attractive features of minimal gauge mediation are not compromised. These values are consistent with the upper bounds on the $\delta_{ij}$ given below \cite{Iyer:2013axa}

 \begin{table}[htbp]
 	\caption{High scale ($M_{GUT}$) $\delta's$ for squark and slepton sector corresponding to the spectrum in Eq.(\ref{spectrum}) }
 	\begin{center}
 		\begin{tabular}{ |c|p{1.5cm}p{1.0cm}|p{1.5cm}p{1.0cm}|p{1.5cm}p{1.0cm}|p{1.5cm}p{1.5cm}p{1.0cm}| }
 			\cline{1-10}
 			(i,j)&$|\delta^Q_{LL}|$ & $|\delta^L_{LL}|$&$|\delta^D_{LR}|$ & $|\delta^U_{LR}|$&$|\delta^D_{RL}|$ & $|\delta^U_{RL}|$&$|\delta^D_{RR}|$ & $|\delta^E_{RR}|$&$|\delta^U_{RR}|$ \\
 			\cline{1-10}
 			12&$1$  &1& $10^{-15}$ &$10^{-10}$&$10^{-16}$&$10^{-8}$& 0.69&1&0.03\\
 			13&$1$  &1& $10^{-11}$ &$10^{-9}$&$10^{-13}$&$10^{-7}$& 1&1&1\\
 			23&$1$  &1& $10^{-8}$ &$10^{-2}$ &$10^{-9}$ &$0.04$& 0.69&1&0.03\\  
 			\cline{1-10}
 		\end{tabular}
 	\end{center}
 	\label{deltahigh}
 \end{table}

 \begin{table}[htbp]
 	\caption{Weak scale $\delta's$ for squark and slepton sector corresponding to the spectrum in Eq.(\ref{spectrum}) }
 	\begin{center}
 		\begin{tabular}{ |c|p{1.5cm}p{1.0cm}|p{1.5cm}p{1.0cm}|p{1.5cm}p{1.0cm}|p{1.5cm}p{1.5cm}p{1.0cm}| }
 			\cline{1-10}
 			(i,j)&$|\delta^Q_{LL}|$ & $|\delta^L_{LL}|$&$|\delta^D_{LR}|$ & $|\delta^U_{LR}|$&$|\delta^D_{RL}|$ & $|\delta^U_{RL}|$&$|\delta^D_{RR}|$ & $|\delta^E_{RR}|$&$|\delta^U_{RR}|$ \\
 			\cline{1-10}
 			12&$10^{-4}$    &$10^{-5}$& $10^{-16}$ &$10^{-13}$&$10^{-14}$&$10^{-9}$& $10^{-6}$&$10^{-3}$&$10^{-8}$\\
 			13&$0.005$    &$10^{-5}$& $10^{-15}$ &$10^{-10}$&$10^{-10}$&$10^{-6}$&$10^{-5}$&$10^{-2}$ & $10^{-4}$\\
 			23&$0.006$    &$10^{-4}$& $10^{-11}$ &$10^{-9}$ &$10^{-9}$ &$10^{-6}$&$10^{-7}$&$10^{-2}$ &$10^{-5}$\\  
 			\cline{1-10}
 		\end{tabular}
 	\end{center}
 	\label{delta}
 \end{table}

\begin{table}
	\caption{Upper bounds on the flavour violating parameter $\delta^{down}$ obtained for $\tilde m_q=2.1$ TeV and $\tilde m_l=0.7$ TeV}
	\begin{center}
		\begin{tabular}{ |c|p{1.5cm}p{1.0cm}|p{1.5cm}p{1.5cm}|p{1.5cm}p{1.5cm}|p{1.5cm}p{1.5cm}| }
			\cline{1-9}
			\cline{1-9}
			(i,j)&$|\delta^Q_{LL}|$ & $|\delta^L_{LL}|$&$|\delta^D_{LR}|$ & $|\delta^E_{LR}|$&$|\delta^D_{RL}|$ & $|\delta^E_{RL}|$&$|\delta^D_{RR}|$ & $|\delta^E_{RR}|$ \\
			\cline{1-9}
			12&0.053 &$0.0002$& $0.0003$ &$3.8\times 10^{-6}$ &$0.0003$ &$3.8\times10^{-6}$&0.03&0.03\\
			13&$0.34$  &0.14& $0.06$ &$0.03$&0.06&$0.03$&0.26& -\\
			23&0.61 &$0.16$&  $0.01$&0.04&$0.02$&0.04&0.84&-\\  
			\cline{1-9}
		\end{tabular}
	\end{center}
	\label{exptconst}
\end{table}
To conclude this section, we point out that it is also possible to generate flavourful soft terms in 4D models.
One example in 4D is to consider Froggatt-Nielsen (FN) type of models 
which have an additional horizontal family symmetry $U(1)_X$ \cite{Froggatt:1978nt}. They were introduced to explain the observed hierarchy in the fermion mass and the mixing angles by assigning
different charges to different generations of fermion multiplets under $U(1)_X$. Requiring the Kahler potential be canonical, rendered a non-trivial flavour structure to the soft
mass sector. One drawback of these models is that $U(1)_X$ group is anomalous and they are only cancelled by the Green-Schwarz anomaly cancellation mechanism \cite{Green:1984sg}.
This severely restricts the parameter space of the FN charges of the multiplets which satisfy the fermion mass fits as well as the anomaly cancellation requirement.
Further these solutions are also constrained by FCNC processes. The Randall-Sundrum framework is a more generalised setup which is equivalent to the setup of $U(1)_{FN}$ but not 
as constraining unless unification conditions are imposed \cite{Dudas}
\section{Conclusions}
Gravitational contributions to supersymmetry breaking are typically suppressed in cases where gauge mediated supersymmetry breaking contributions
are dominant.  In the present work, we show that this is not the case if the supersymmetry breaking happens in multiple sectors which are
parametrised in terms of different spurion fields.  If they are hierarchical, one can have a situation where gravity contributions can play a supporting 
role to the gauge mediated contributions and help in generating the right Higgs mass.  Of course, this would require that the gravitational contribution
should be flavourful which we have achieved by using a Randall Sundrum set up.  In the course of this work, we have presented simple ``rules of thumb"
 to RGE equations when supersymmetry is broken by multiple sectors at multiple scales. 

 \appendix
 \section{Analytic expressions for soft masses in models with two sectors of SUSY breaking}
 \label{appendix1}
 In this section we give the analytic expressions for the soft masses at the weak scale due to SUSY breaking from two different hidden sectors. 
 To begin with, consider the gauginos and the A terms. The net contribution at any scale due to the presence of both gauge and gravity sectors is just the linear sum of 
 individual contributions due to pure gauge and pure gravity mediation. 
 We define the following useful functions:
  \begin{eqnarray}
 z_i(t)&=&{1\over (1+b_i\,\tilde{\alpha}_i(0)\,t)}\;\;\;\;\; z'_i(t)={1\over (1+b_i\,\tilde{\alpha}_i(t_g)\,t)}\nonumber\\
 k_i(t)&=&\frac{1}{2b_i}(1-z_i(t)^2)\;\;\;\;\;k'_i(t)=\frac{1}{2b_i}(1-z'_i(t)^2)
 \end{eqnarray}
 where 
 \begin{eqnarray}
  t_g&=& 2\, Log\left({M_{X_2}\over M_{X_1}}\right)\;\;\;\;t_z= 2\, Log\left({M_{X_2}\over m_{susy}}\right)\nonumber\\
 \end{eqnarray}
and $\tilde{\alpha}_i(t)= z_i(t)\,\tilde{\alpha}_i(0)$ and $\tilde{\alpha}_i(0)$ is the value at $M_{X_2}$.
 Without loss of generality we can assume $m_{susy}<M_{X_1}<M_{X_2}$. 
 
 The solutions to the one loop renormalization group evolution equations (RGE) for the gauginos can be 
 solved exactly and are given as
 \begin{eqnarray}
  \tilde M_i(m_{susy}) = M_iz_i(t_z)+M^{\prime}_iz'_i(t_z) \nonumber\\
  \end{eqnarray}
where $M_a(M'_a)$ are the boundary values for the gauginos at scale $M_{X_2}(M_{X_1})$. In our notation quantities with $\tilde{}$~  represent the expressions at the $m_{susy}$ scale
due two sectors of SUSY breaking.
 
Similarly the $A$ terms also add up linearly and are given as \footnote{The semi-analytic expressions are for $M_{X_1}=10^6$ and $M_{X_2}=M_{Pl}$}
\begin{eqnarray}
 \tilde A&=& A(t_z)+A'(t_{z})\nonumber\\
  A(t_z)&=&0.33A_0 - 0.03 M_1 - 0.26 M_2 - 1.8 M_3\nonumber\\
 A'(t_z)&=& -0.002\Lambda
\end{eqnarray}
where $A_0$ is the boundary value for the trilinear coupling due to gravity mediation.

 Next we consider the lighter generations.
 In the limit the Yukawa coupling $Y_{b,\tau}\rightarrow 0$, the solutions to the RGE equations for the first two generation squarks and the sleptons take a very simple analytic form.
The solutions for the squarks and $\tilde m^2_{H_d}$ except $\tilde m^2_{Q_3,U_3,H_u}$ is given as 
\begin{eqnarray}
 \tilde m^2_{Q_{1,2}}&=&m^2_Q(t_z)+{m}^{\prime 2}_{Q_{1,2}}(t_z)+
\frac{32}{3}M'_{3}M_3z_3(t_g)k'_3(t_z)+ 6M'_{2}M_2z_2(t_g)k'_2(t_z)+\frac{2}{15}M'_{1}M_1z_1(t_g)k_1'(t_z)\nonumber\\
 \tilde m^2_{U_{1,2}}&=&m^2_{U_{1,2}}(t_z)+{m}^{\prime 2}_{U_{1,2}}(t_z)+\frac{32}{3}M'_{3}M_3z_3(t_g)k'_3(t_z)+\frac{32}{15}M'_{1}M_1z_1(t_g)k'_1(t_z)\nonumber\\
\tilde m^2_{D_{1,2,3}}&=&m^2_{D_{1,2,3}}(t_z)+{m}^{\prime 2}_{D_{1,2,3}}(tgz)+\frac{32}{3}M'_{3}M_3z_3(t_g)k'_3(t)+\frac{8}{15}M'_{1}M_1z_1(t_g)k_1'(t)\nonumber\\
\tilde m^2_{H_d}&=&m^2_{H_d}(t_z)+{m}^{\prime 2}_{H_d}(t_z)+6M'_{2}M_2z_2(t_g)k'_3(t_z)+\frac{6}{5}M'_{1}M_1z_1(t_g)k'_1(t_z)
\label{squark12}
\end{eqnarray}

The corresponding equations for the sleptons can be similarly written as
\begin{eqnarray}
 \tilde m^2_{L}&=&m^2_L(t_{z})+{m}^{\prime 2}_{L}(t_z)+6M'_{2}M_2z_2(t_g)k'_2(t_z)+\frac{6}{5}M'_{1}M_1z_1(t_g)k_1'(t_z)\nonumber\\
\tilde m^2_{E}&=&m^2_E(t_{z})+{m}^{\prime 2}_{E}(t_z)+\frac{24}{15}M'_{1}M_1z_1(t_g)k_1'(t_z)\nonumber\\
\end{eqnarray}

\subsection{Semi-analytic expression for $m^2_{Q_3,U_3,H_u}$}
The expressions for $m^2_{Q_3,U_3,H_u}$ are more complicated as it involves the top quark Yukawa coupling which cannot be set equal to zero.
However they follow the generic form of Eq.(\ref{softhybrid}).
We present semi-analytic expressions for them as follows:
\begin{eqnarray}
 \tilde m^2_{Q_3}&=&m^2_{Q_3}(t_z)+m^{\prime 2}_{Q_3}(t_z)+0.0001A_0\Lambda+0.0054M_3\Lambda\nonumber\\
 \tilde m^2_{U_3}&=&m^2_{U_3}(t_z)+m^{\prime 2}_{U_3}(t_z)+0.0004A_0\Lambda+0.0045M_3\Lambda\nonumber\\
 \tilde m^2_{H_u}&=&m^2_{H_u}(t_z)+m^{\prime 2}_{H_u}(t_z)+0.0006A_0\Lambda-0.0041M_3\Lambda
\end{eqnarray}
where $\Lambda=\frac{\langle F_1 \rangle}{M_{X_1}}$. 
In the above expression we have neglected the terms proportional to $M_1$ and $M_2$ as their co-efficients are small compared to that of $M_3$. 
The semi-analytic expressions for the soft terms at $m_{susy}$ scale due to pure gravity mediation is given as
\begin{eqnarray}
 m^2_{Q_3}&=&-0.0369233 A_0^2 + 0.00443031 A_0 M_1 + 0.00154387 M_1^2 + 
 0.0266445 A_0 M_2 - 0.00205633 M_1 M_2\nonumber\\ &+& 0.233435 M_2^2 + 0.1096 A_0 M_3 - 
 0.00916469 M_1 M_3 - 0.0610636 M_2 M_3 + 3.31617 M_3^2 \nonumber\\&+& 0.3312 m_{Q_3}^{(0)2} - 
 0.11145 (m_{Hu}^{(0)2} - 5 m_{Q_3}^{(0)2} + m_{U_3}^{(0)2})\nonumber\\
 m^2_{U_3}&=& -0.0738465 A_0^2 + 0.00886063 A_0 M_1 + 0.0287372 M_1^2 + 
 0.0532891 A_0 M_2 - 0.00411266 M_1 M_2\nonumber\\ &-& 0.0134655 M_2^2 + 0.2192 A_0 M_3 - 
 0.0183294 M_1 M_3 - 0.122127 M_2 M_3 + 3.16522 M_3^2\nonumber\\ &-& 
 0.222901 (m_{Hu}^{(0)2} + m_{Q_3}^{(0)2} - 2 m_{U_3}^{(0)2}) + 0.3312 m_{U_3}^{(0)2}\nonumber\\
 m^2_{H_u}&=&-0.11077 A_0^2 + 0.0132909 A_0 M_1 + 0.0156242 M_1^2 + 0.0799336 A_0 M_2 - 
 0.00616898 M_1 M_2\nonumber\\ &+& 0.21997 M_2^2 + 0.3288 A_0 M_3 - 0.0274941 M_1 M_3 - 
 0.183191 M_2 M_3 - 0.452999 M_3^2 \nonumber\\ &+& 0.3312 m_{Hu}^{(0)2} - 
 0.334351 (-m_{Hu}^{(0)2} + m_{Q_3}^{(0)2} + m_{U_3}^{(0)2})
 \end{eqnarray}
where $m^{(0)}_{Q_3,U_3,Hu}$ represents the boundary value of the soft masses due to gravity mediation. The corresponding expressions due to gauge mediation with messenger scale at $M_{X_1}=10^6$
is given as

 \begin{eqnarray}
m^{\prime 2}_{Q_3} &\approx& 1\times10^{-4}\Lambda^2\nonumber \\
 m^{\prime 2}_{U_3} &\approx& 7.62\times10^{-5}\Lambda^2\nonumber\\
 m^{\prime 2}_{H_u} &\approx& -4.65\times10^{-5}\Lambda^2
\end{eqnarray}
.

\bibliographystyle{ieeetr}
\bibliography{rescuegauge.bib}
\end{document}